\documentclass[reprint,aps,prl,superscriptaddress,amsmath,amssymb,bibnotes,showkeys]{revtex4-1}
\usepackage{graphicx}

\usepackage{graphicx}
\usepackage{amsmath}
\usepackage{dcolumn}
\usepackage{bm}
\usepackage{braket} 
\usepackage{bbold}
\usepackage{diagbox}
\usepackage{multirow}
\usepackage{lpic} 
\usepackage{url}
\usepackage{array}
\usepackage{color}
\usepackage[
 colorlinks=true,
 urlcolor=blue,
 citecolor=blue,
 linkcolor=blue,
 bookmarks=false,
 pdfstartview={FitH},
]{hyperref}

\def\cm{cm$^{-1}$}
\def\CaK{CaKFe$_4$As$_4$}
\def\A{A$_{1g}$}
\def\B{B$_{1g}$}
\def\BB{B$_{2g}$}
\def\Tc{$T_c$}

\begin{document}
\title{High \Tc\, superconductivity in \CaK\, in absence of nematic 
fluctuations}

\author{W.-L.~Zhang}
\email{wz131@physics.rutgers.edu}
\affiliation{Department of Physics $\&$ Astronomy, Rutgers
University, Piscataway, New Jersey 08854, USA} 
\author{W. R. Meier}
\affiliation{Department of Physics and Astronomy, Iowa State
University, Ames, Iowa 50011, USA}
\affiliation{Division of Materials Science and Engineering, Ames
Laboratory, Ames, Iowa 50011, USA} 
\author{T. Kong}
\affiliation{Department of Physics and Astronomy, Iowa State
University, Ames, Iowa 50011, USA}
\affiliation{Division of Materials Science and Engineering, Ames
Laboratory, Ames, Iowa 50011, USA} 
\author{P. C. Canfield}
\affiliation{Department of Physics and Astronomy, Iowa State
University, Ames, Iowa 50011, USA}
\affiliation{Division of Materials Science and Engineering, Ames
Laboratory, Ames, Iowa 50011, USA} 
\author{G.~Blumberg}
\email{girsh@physics.rutgers.edu}
\affiliation{Department of Physics $\&$ Astronomy, Rutgers
University, Piscataway, New Jersey 08854, USA}
\affiliation{National Institute of Chemical Physics and Biophysics,
Akadeemia tee 23, 12618 Tallinn, Estonia}

\date{\today}

\begin{abstract}

We employ polarization-resolved Raman spectroscopy to study
multi-band stoichiometric superconductor \CaK. 
The \BB\, symmetry Raman response shows no signatures of 
Pomeranchuk-like electronic nematic fluctuations which is observed 
for many other Fe-based superconductors.  
In the superconducting state, we identify three pair-breaking peaks at 
13.8, 16.9 and 21~meV and full spectral weight 
suppression at low energies.  
The pair-breaking peak energies in Raman response are about 20\%
lower than twice the gap energies as measured by single-particle
spectroscopy, implying a sub-dominant $d$-wave symmetry interaction. 
We analyze the superconductivity induced phonon self-energy effects and give an
estimation of weak electron-phonon coupling constant $\lambda^\Gamma$=0.0015.
\end{abstract}

\maketitle

\textit{Introduction}
-- Understanding the pairing mechanism in the Fe-based
superconductors (FeSC) remains focused topic of research not only due 
to a high superconducting transition  
temperature \Tc, but also because of the unusual  
properties and interplay with other electronic 
degrees of freedom, such as nematicity and 
magnetism~\cite{Paglione_nphys2010,Wang_science2011,Chubukov2012,Fernandes_NatPhys2014,Hirschfeld_CRP2016,QSi_natreview2016}. 
For FeSCs, superconductivity usually appears in close proximity to nematic 
and/or magnetic ground states~\cite{Shibauchi2014,Fernandes_NatPhys2014}. 

FeSCs are believed to be unconventional superconductors because
the pairing mechanism is due to electronic rather than phononic interactions. 
It has been proposed that
spin-fluctuation can lead to s$_\pm$ symmetry pairing with
sign reversal gap function between Fermi surface (FS) pockets around $\Gamma$ and M
point~\cite{Mazin_PRL2008,Hirschfeld_RRP2011}. 
In addition, strong
electronic nematic fluctuations have been observed for the materials
with highest \Tc~\cite{Fisher_Science2012,Gallais_PRL2013,Kuo_Science2016} and, 
hence, a mechanism for \Tc\, enhancement by nematic fluctuations has been 
proposed~\cite{Kivelson_PRL2015}. 
As the alternatives, orbital-fluctuation in the presents of strong 
electron-phonon interaction were considered to be another possible 
pairing glue which could lead to nodeless $s_{++}$ symmetry
pairing~\cite{Kontani_PRL2010,WeiKu_PRL2010,Saito_PRB2010}. 
Raman scattering has been used to study the spectroscopy of superconducting (SC)  
gap~\cite{Klein_PhysRevB1984,Devereaux_RMP2007,Thorsmolle_PRB2016,Gallais_PRL2016,SWu_PRB2017,Bohm_Arxiv1703}, 
the dynamics of electronic nematic 
fluctuations~\cite{Gallais_PRL2013,Hackl_nphys2016,Thorsmolle_PRB2016,Wu_2017a}, 
as well as the strength of electron-phonon
interactions~\cite{Cooper_PRB1988,Thomsen_PRB1988,Zeyher1990,Mialitsin_PRB2007}. 

Recent discovery of a new class of stoichiometric and strictly 
tetragonal FeSCs
Ca$A$Fe$_4$As$_4$ ($A$ = K, Rb, Cs) with rather high \Tc\,\,(31-36~K) provides an ideal platform for spectroscopic investigation of
superconductors in clean limit, to decide on the role of 
nematicity for high \Tc\, superconductivity in FeSCs~\cite{Meier_PRB2016,Akira_JACS2016,Meier_PRM2017,XGQiu_PRB2017,Iyo_2018}. 
Electron doping \emph{via} substitution of Co or Ni for Fe, and/or  
application of pressure, suppresses \Tc\, and induces an exotic 
spin-vortex crystal order, while the system remains
tetragonal even for the magnetically ordered phase~\cite{Meier_nqm2018,Canfield_PRB2017,Canfield_1803}.

\begin{figure}[!]
\includegraphics[width=\columnwidth]{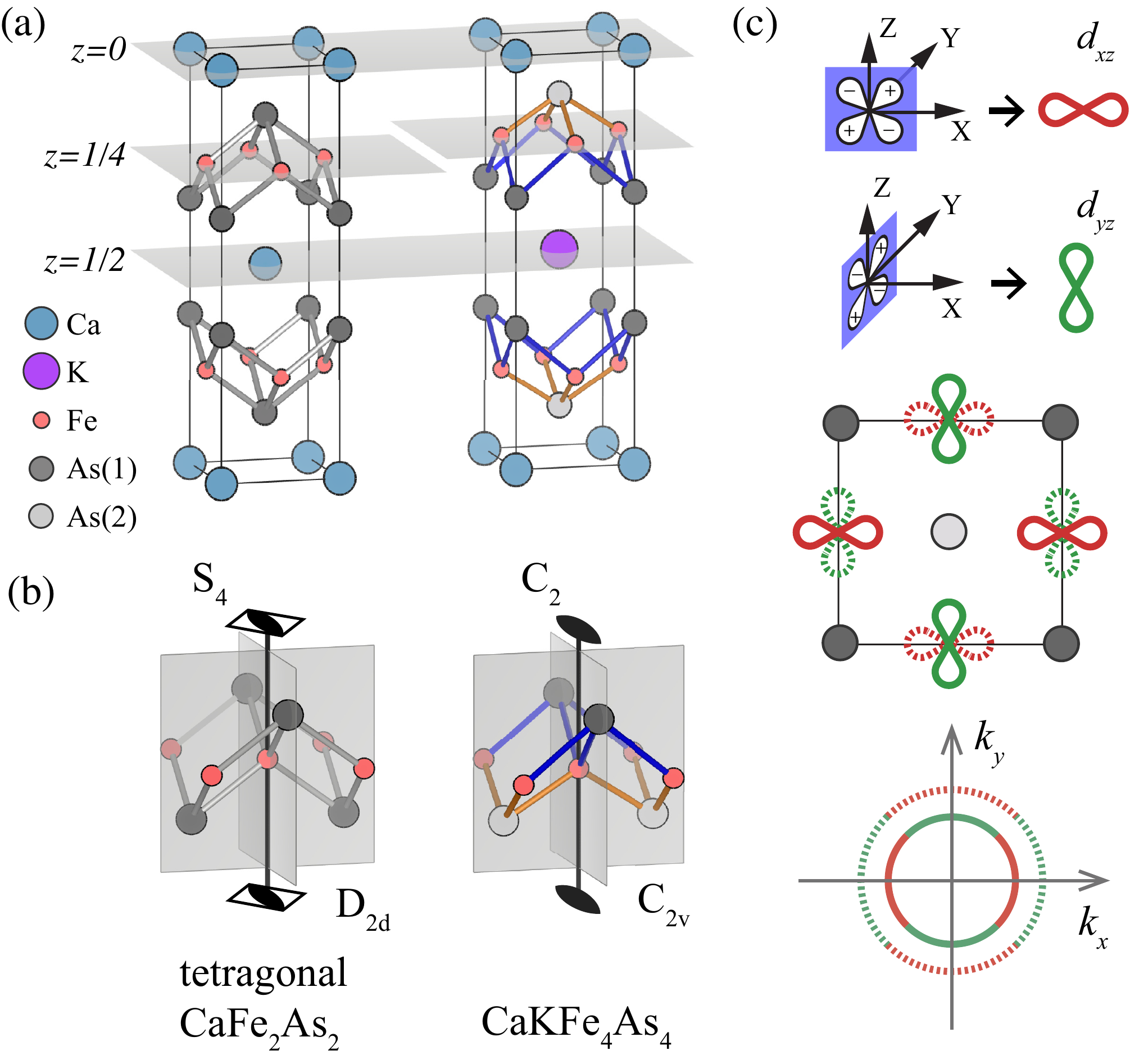}
\caption{\label{Fig0} 
(a) The comparison between CaFe$_2$As$_2$ and
\CaK\, lattice structures.
For \CaK, the Fe sites are shifted away from
the high symmetry z=1/4 and 3/4 planes, causing two distinct As-Fe bond
distances (shown in blue and orange). 
(b) The reduction of the Fe site-symmetry from $D_{2d}$ for 
CaFe$_2$As$_2$ to $C_{2v}$ for \CaK. 
(c) For \CaK, 
a sketch of the partially occupied $d_{xz}/d_{yz}$ orbital order 
(the upper panel for real space), 
and derived two FS pockets in the vicinity of the $\Gamma$ point (the 
lower panel for momentum space). 
Solid and dotted lines denote differences in the orbital occupation 
which induce a static quadrupole moment on the Fe sites
with a checkerboard order. 
}
\vspace{-5mm}
\end{figure}

In this Letter, we report polarization-resolved Raman spectroscopic
study of \CaK\, single-crystals. 
In contrast to the data from many FeSCs for which strong Pomeranchuk-like 
electronic nematic fluctuations give rise to intense electronic Raman 
continuum in the XY-quadrupole symmetry channel, 
the electronic Raman signal for \CaK\, is weak and isotropic. 
Below \Tc, we observe complete suppression of the
low-frequency spectral intensity and development of three SC
pair-breaking coherence peaks in the XY symmetry electronic Raman
response, which implies that all FS pockets are fully
gapped. 
The pair-breaking peak energies $2\Delta$ = 13.8, 16.9 and 21~meV are 
about 20\% lower
than twice the gap energies measured by single-particle spectroscopies.
We attribute such renormalization to
finite-state interactions in sub-dominant $d$-wave SC pairing channel.
We also observe superconductivity induced phonon self-energy effect
and estimate the electron-phonon (e-p) coupling to be weak, with
coupling constant $\lambda^\Gamma$ = 0.0015.


\textit{Crystallographic structure}
-- The \CaK\,\,single crystals (\Tc\,\,= 35 K) used in this study
were synthesized by flux method~\cite{Meier_PRB2016,Meier_PRM2017}. 
The structure may be considered as a modification to the extensively
studied tetragonal 122 FeSCs with the body-centered structure
$I4/mmm$.
For \CaK, every other plane of Ca atom is replaced by
K, reducing the crystallographic space group to primitive $P4/mmm$
(point group $D_{4h}$) and therefore doubling the number of atoms in
the primitive cell. 
In Figs.~\ref{Fig0}(a-b) we compare the tetragonal CaFe$_2$As$_2$
and \CaK\,\,lattices.

For the body-centered CaFe$_2$As$_2$, the Fe layers are at the high
symmetry z=1/4 and 3/4 planes, and each Fe has $D_{2d}$ site symmetry
with $S_4$ axis along z-direction, which imposes the degeneracy of Fe
3$d_{xz}$ and 3$d_{yz}$ orbitals. 
Such orbital degeneracy is a common feature for most of the
FeSCs~\cite{Vafek_PRB2013, JPHu_PRX2012}. 
It has been demonstrated that effect of partially filled 
$d_{xz}$/$d_{yz}$ orbital degeneracy is significant:
it causes dynamical charge oscillations between quasi-degenerate orbitals in 
sub-THz frequency range. 
This gives rise to fluctuating charge
ferro-quadrupole moment with an amplitude proportional to the local
oscillating charge imbalance
$n_{xz}-n_{yz}$~\cite{Thorsmolle_PRB2016}. 
The soft ferro-quadrupole fluctuations often show critical behavior 
leading to a $d$-wave Pomeranchuk
instability~\cite{Pomeranchuk_1958,Thorsmolle_PRB2016}. 
These fluctuations most dramatically manifest themselves in the 
low-frequency part of $XY$-symmetry Raman response as an 
overdamped quasi-elastic feature in the normal
state~\cite{Gallais_PRL2013,Thorsmolle_PRB2016,Hackl_nphys2016} which 
undergoes a metamorphosis into a coherent in-gap collective mode
below
\Tc\,~\cite{Khodas_PRB2014,Thorsmolle_PRB2016,Gallais_PRL2016,SWu_PRB2017,Maiti_PRB2017}.

In contrast, for \CaK\,\,structure, the ordered alternating Ca and K
layers above
and below Fe-As layers cause the Fe positions shift away from the
high symmetry planes~\cite{Eremin_PRB2017,Furukawa_PRB2017}. 
The shift gives 
rise to nonequivalent Fe-As bond
distances for the As atoms above and below the Fe layer, thus,
reducing the Fe site symmetry to $C_{2v}$ (Fig.~\ref{Fig0}(a-b)). 
The removal of the
$S_4$ symmetry on the Fe sites takes away the degeneracy of the 
partially occupied Fe
$d_{xz}$/$d_{yz}$ orbitals~\cite{Eremin_PRB2017}, giving rise to a 
static charge imbalance between these two orbitals, and, hence, 
creates a static $d_{x^{2}-y^{2}}$-symmetry quadrupole moment on each
Fe site.
Because the orbital character of the lower energy state flips between 
two neighboring Fe sites, the structure forms a static checkerboard 
anti-quadrupole order (Fig.~\ref{Fig0}(c)). 
The stiffness of the static anti-quadrupole order parameter precludes 
Pomeranchuk-like fluctuations for the \CaK\,\,compound.


\begin{figure}[!]
\includegraphics[width=0.85\columnwidth]{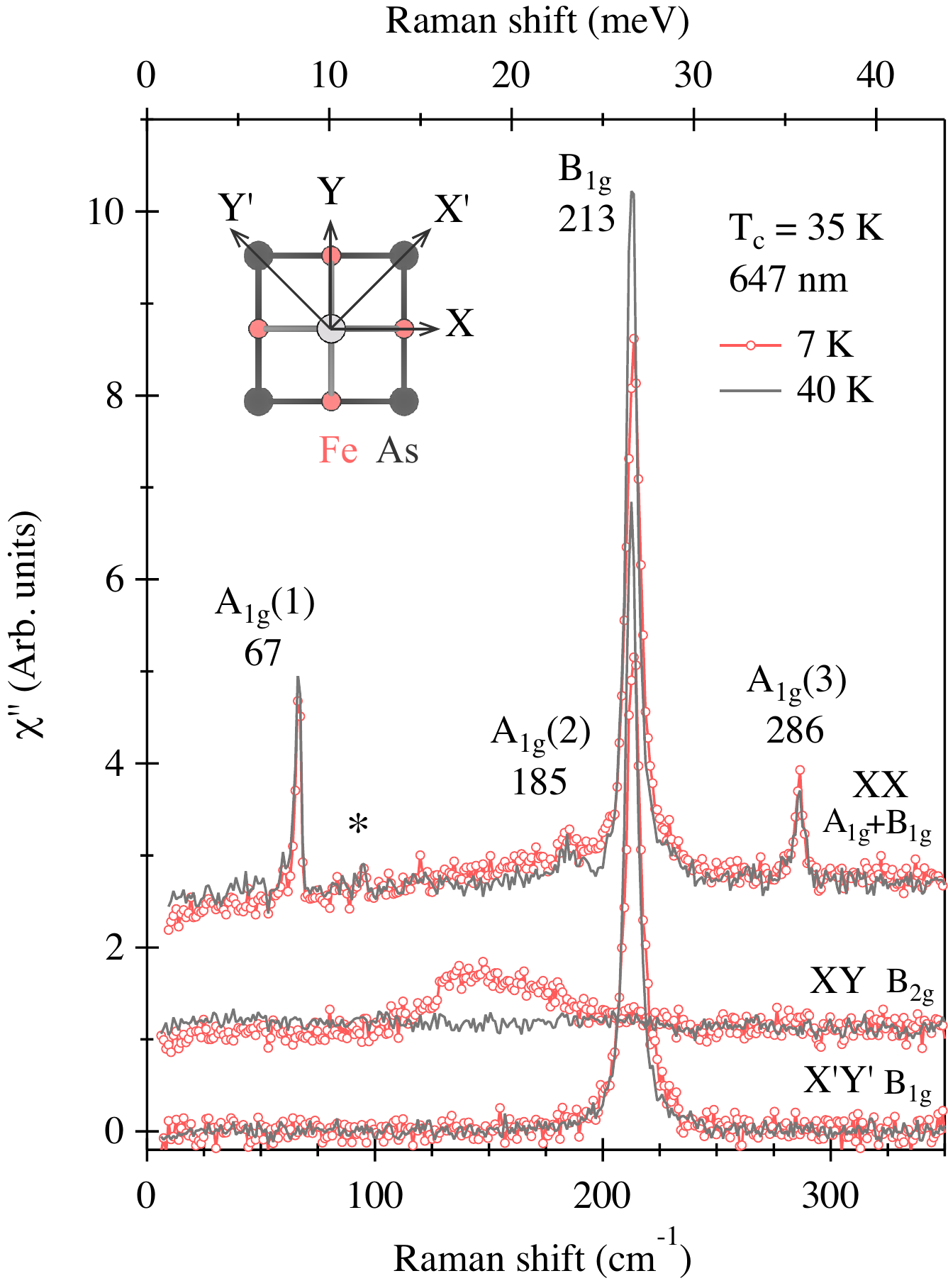}
\caption{\label{Fig1} Raman spectra for single crystal \CaK\,\,at 40
and 7~K. The XY and XX polarized spectra are offset by 1 and 2 units,
respectively. 
Inset: top view of the Fe-As layer and axes notation.}
\vspace{-3mm}
\end{figure}


\textit{Experimental}
-- Polarization-resolved Raman scattering measurements were performed
in quasi-back scattering geometry from the natural cleaved (001)
surface. 
We use $\mu\nu=$ XX, XY and
X$^\prime$Y$^\prime$ scattering geometries where $\mu\nu$ is short
for $\bar{\mathrm{Z}}$($\mu\nu$)Z in Porto's notation.
These geometries allow coupling to excitations with the
\A+\B\,\,symmetry by XX polarization, \BB\,\,symmetry by XY
polarization, and \B\,\,symmetry by X$^\prime$Y$^\prime$
polarization. 

The crystals were loaded into a continuous helium flow optical
cryostat immediately after being cleaved in a nitrogen gas filled
glove bag connected to the cryostat.
We used the 647 nm line of a Kr$^+$ laser, where the laser beam was
focused to a 50 $\times$ 50 $\mu$m spot. 
The laser power was kept below 10 mW in the normal state and 2.3 mW
for the SC state to reduce laser heating. 
The temperatures were corrected for the laser heating. 

The Raman signal was collected and analyzed by a triple-grating 
spectrometer with 1.5~\cm\,\,spectral resolution. 
All spectra were corrected for spectral response and background
determined from the X$^\prime$Y$^\prime$ symmetry electronic
continuum to obtain the Raman scattering intensity 
$I_{\mu\nu}(\omega, T)$~\footnote{The background is determined from
the
X$^\prime$Y$^\prime$ symmetry scattering intensity after subtracting
the \B\,\,phonon, see supplement I in~\cite{WZhang_FeSe2017}}.
The Raman response function $\chi^{\prime\prime}_{\mu\nu}(\omega, T)$
is related to $I_{\mu\nu}(\omega, T)$ by the Bose distribution factor
$n(\omega, T)$: $I_{\mu\nu}(\omega, T)$ = [1+$n(\omega, T)$]
$\chi^{\prime\prime}_{\mu\nu}(\omega, T)$.

\begin{figure}[!t]
\includegraphics[width=\columnwidth]{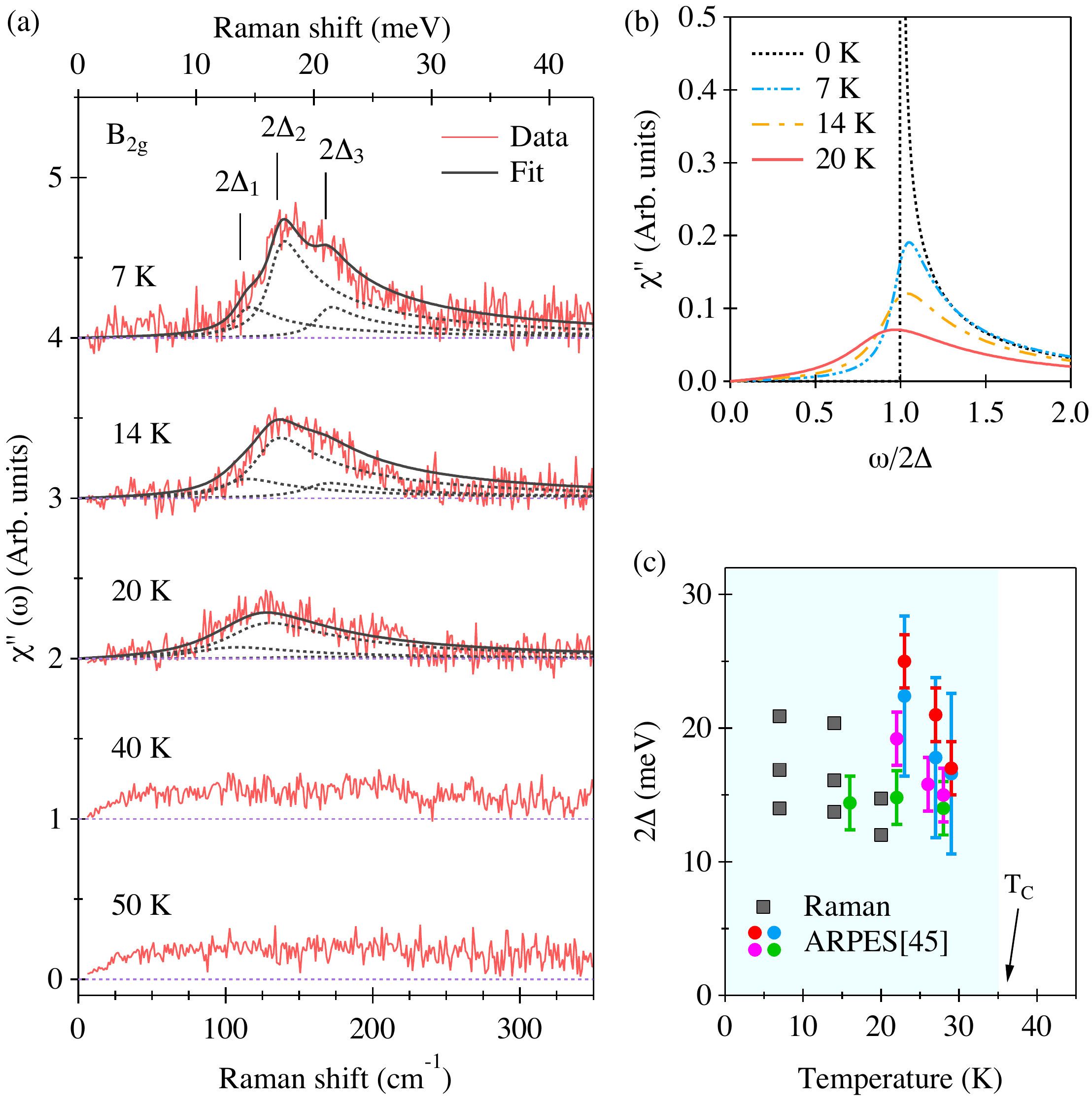}
\caption{\label{Fig2} (a) Temperature evolution of the
\BB\,\,symmetry Raman response. The black solid lines are fits to the
data. Below \Tc, the data are fit with three BCS gap response at
finite temperature. (b) Temperature evolution of Raman response for a
BCS gap. (c) The SC gap magnitude $2\Delta(T)$ measured by Raman
spectroscopy and ARPES~\cite{Mou_PRL2016}. }
\end{figure}

\textit{Raman continuum}
-- In Fig.~\ref{Fig1} we show Raman spectra at temperatures above and
below \Tc\,\,for XX, XY and X$^\prime$Y$^\prime$ polarizations. 
Above \Tc, besides four sharp phonon modes, we observe a weak and 
featureless electronic continuum for all polarizations. 
Below \Tc, for the XY scattering geometry, a broad feature appears at 
around 150~\cm, 
which is an indication of the SC gaps formation. 
For the XX and
X$^\prime$Y$^\prime$ scattering geometries, the electronic continuum
shows little change across \Tc. 
In the following we focus on the electronic Raman response in the
\BB\,\,(XY) symmetry channel. 

The common signature for most of the FeSCs is strong XY-symmetry 
Raman response. 
The enhancement of the Raman susceptibility has been 
attributed to the $d_{XY}$-symmetry Pomeranchuk-like fluctuations
with 
Drude-like low-frequency response in the normal state and sharp
in-gap 
collective mode in the superconducting state, both arising from
strong 
oscillating charge ferro-quadrupole moments on the Fe sites due to 
the $d_{xz}$/$d_{yz}$ orbital degeneracy~\cite{Thorsmolle_PRB2016,
Yamase_PRB2004,Yamase2011,Yamase_PRB2013}. 
For \CaK\,\,structure this orbital degeneracy is lifted, and because 
of that, the XY-symmetry Raman susceptibility is quite weak and 
featureless. 

\textit{SC pair-breaking peaks} -- 
Below \Tc, we observe a suppression of the low-frequency spectral
weight and development of a feature between 10 to 30 meV. Upon
cooling, the intensity and central frequency of this feature
increase. At the lowest measured temperature (7~K), the spectral
intensity below 12~meV is completely suppressed, indicating that the
SC gap is nodeless. 
We detect no SC coherence peak for a gap near 5 meV, as reported in 
Refs.~\cite{Hillier_PRB2017,Prozorov_PRB2017}. 
We note that the spectra does not display an in-gap collective mode 
which is a common feature for many FeSCs that show Pomeranchuk-like 
nematic fluctuations above 
\Tc~\cite{Khodas_PRB2014,Thorsmolle_PRB2016,Gallais_PRL2016,SWu_PRB2017}.

We fit the data with three-gap model, 
$\chi^{\prime\prime}=\sum\limits_{i=1}^3\alpha_i\chi_i^{\prime\prime}$.
Here $\alpha_i$ denotes the spectral weight, and
$\chi_i^{\prime\prime}(\omega, \Delta_i, T)
=4\Delta_i^2/\omega\sqrt{\omega^2-4\Delta_i^2} \ast L(\omega, T)$
denotes
BCS coherence peaks with SC gap energies 
$2\Delta_i$~\cite{Klein_PhysRevB1984} 
convoluted with temperature dependent 
Lorentzian~\cite{Blumberg_200775}. 
The effects of temperature broadening are depicted in
Fig.~\ref{Fig2}(b).

In Fig.~\ref{Fig2}(c) we summarize the temperature dependence of the
three SC gap amplitudes $2\Delta_i(T)$. 
At 7 K, the gaps energies 110~\cm (13.8 meV), 135~\cm (16.9 meV), and
168~\cm (21 meV) correspond to the 2$\Delta/k_B$\Tc= 6.9, 5.5, 
and 4.5, ratios respectively. 
It appears that the pair-breaking peak energies measured by Raman
are consistently near 20\% lower than the gap energies determined
for similar temperatures by ARPES~\cite{Mou_PRL2016}. 
We attribute this renormalization of the peak energies to the
finite-state interaction effects in the sub-dominant $d_{xy}$-wave
pairing channel~\cite{Klein_PhysRevB1984, Khodas_PRB2014,
Maiti_PRB2017}.


\begin{table}[!b]
\vspace{-5mm}
\caption{Summary of Raman active phonons mode energies and atomic
displacements for \CaK.}
\label{Table2}
\vspace{-5mm}
\begin{center}
\begin{tabular}{m{6em}m{8em}m{12em}}
\hline\hline
Symmetry&Energy at 40 K&Atomic displacements\\
\hline
$A_{1g}(1)$	& 67 \cm&Fe(z)+As(1)(z)+As(2)(z)\\
$A_{1g}(2)$	& 185 \cm&Fe(z)+As(1)(z)+As(2)(z)\\
$A_{1g}(3)$	&286 \cm&Fe(z)+As(1)(z)+As(2)(z)\\
$B_{1g}(1)$	&213 \cm&Fe(z)\\
\hline
\hline
\end{tabular}
\end{center}
\vspace{-3mm}
\end{table}%
\textit{Phonon self-energy effects}
-- \CaK\,\,crystal (point group $D_{4h}$) contains 10 atoms in a
primitive cell. 
For phonons at the $\Gamma$ point, group-theoretical symmetry decomposition
yields $A_u$+$E_u$ acoustic modes, $4A_{2u}+5E_u$ infrared
active
modes, $3A_{1g}+B_{1g}+4E_g$ Raman active modes, and a B$_{2u}$
silent mode. 
For the scattering experiments from (001) surface, we observe all
three
\A\,\,and a \B\,\,phonons (Fig.~\ref{Fig1}). The phonon energies
and atomic displacements are summarized in
Table~\ref{Table2}.

\begin{figure}[!t]
\includegraphics[width=\columnwidth]{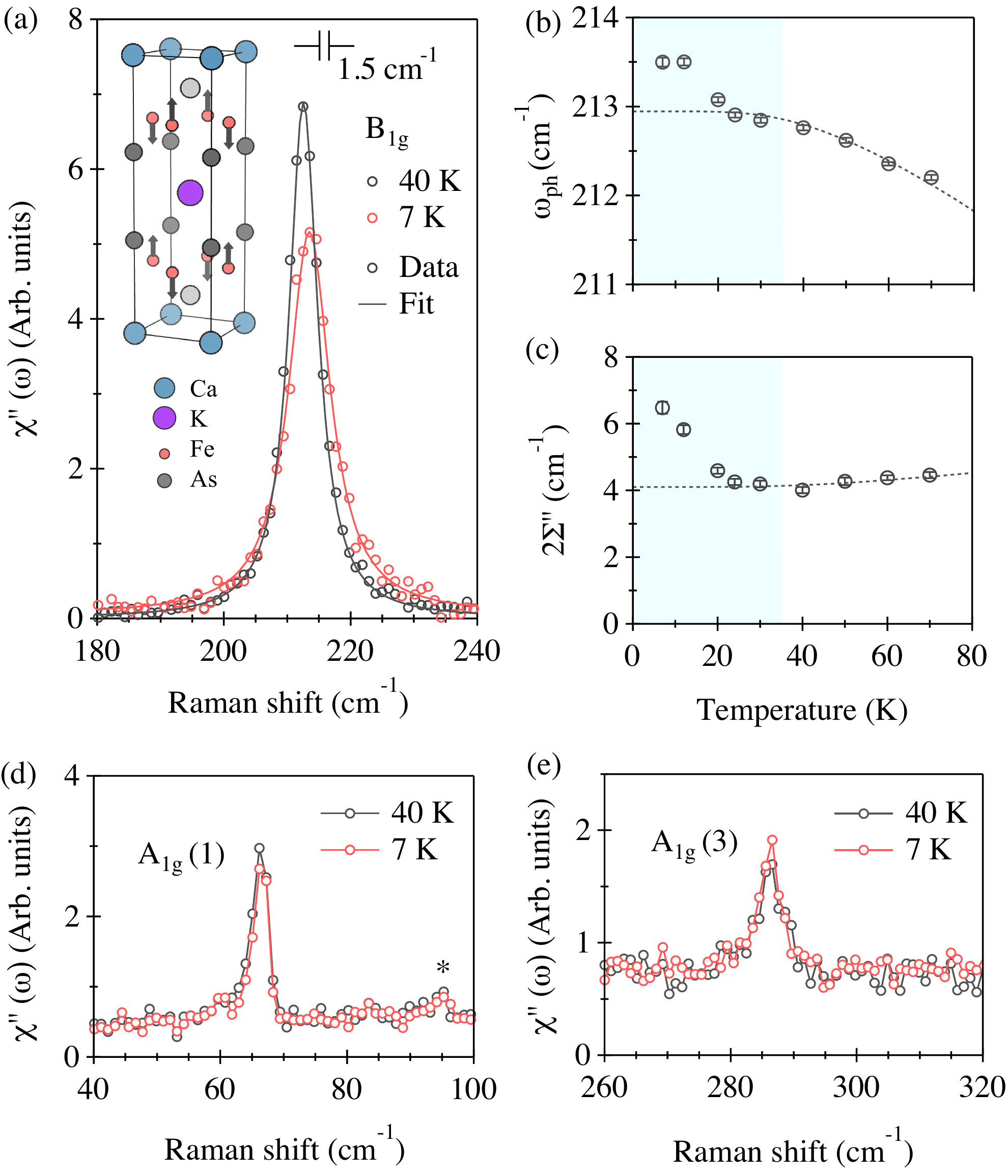}
\caption{\label{Fig4} Temperature evolution of phonon spectra. (a)
The \B\,\,phonon spectra and the fitting curve for above (40 K) and
below (7 K) \Tc. Inset of (a): Atomic displacements of the \B\,\,
phonon. (b)-(c) Temperature dependence of \B\,\,phonon energy
$\omega_{ph}$ and line width $2\Sigma^{\prime\prime}$.
$\Sigma^{\prime\prime}$ has been corrected for the spectrometer
resolution as shown in (a). Dashed lines describe the anharmonic
decay into two phonons obtained from the fit for the normal state.
(d)-(e) The \A(1) and \A(3) phonon modes at 40 and 7 K.
}
\end{figure}

Above \Tc\,\,all modes exhibit a conventional temperature dependence: 
hardening and narrowing upon cooling due to 
anharmonic decay~\cite{Klemens_PR1966,Cardona_PRB1984,CaKfootnote}. 
However, we note that the behavior for the \B\,\,phonon mode in the 
SC state is anomalous: the mode's 
energy and line width increase upon cooling, as shown in
Fig.~\ref{Fig4}(a).

Similar phonon anomalies upon entering into SC state were
reported for MgB$_2$~\cite{Mialitsin_PRB2007} and for cuprate
superconductors~\cite{Cooper_PRB1988,Thomsen_PRB1988,Friedl_PRL1990,McCarty_PRB1991,Blumberg1994,Thomsen_1998}.
The behavior was explained by Zeyher-Zwicknagl's
model~\cite{Zeyher1990} which implies that in the presents of
electron-phonon coupling, phonon self-energy is upward renormalized
when the SC gap opens and the electronic density-of-states is pushed
to the proximity of the phononic mode above the gaps
energies.

It is interesting to note that in contrast to the 67~\cm\,\,\A\,\,phonon (Fig.\ref{Fig1}) that exhibits an asymmetric Fano line shape,
the \B\,\,phonon 
shows nearly perfect Lorentzian shape. 
We attribute this to weak Raman coupling to the
\B\, symmetry electronic continuum: the e-p interaction only
renormalizes the phonon self-energy without showing a Fano
interference in the spectra.

To quantify the superconductivity induced self-energy effects and the 
e-p interaction strength, we fit the data with
$\chi_{ph}^{\prime\prime}(\omega)\propto4\omega_0\Sigma^{\prime\prime}[(\omega^2-\omega_0^2-2\omega_0\Sigma^\prime)^2+4(\omega_0\Sigma^{\prime\prime})^2]^{-1}$,
where $\omega_0$ is the bare phonon frequency, and
$\Sigma=\Sigma^\prime+i\Sigma^{\prime\prime}$ is complex phonon
self-energy~\cite{Zeyher1990}.
Thus, if $\Sigma$ is small, the mode appears at
$\omega_{ph}=\sqrt{\omega_0^2+2\omega_0\Sigma^\prime}$ frequency. 
The fitting results are displayed in Figs.~\ref{Fig4}
(b-c)~\cite{CaKfootnote}.

We calculate the coupling constant $\lambda_{B_{1g}}^\Gamma$ around
$\Gamma$ point~\cite{Rodriguez_PRB1990}: $\lambda=-\kappa\sin u/u$,
where $\kappa=[(\Sigma^\prime(7 K)-\Sigma^\prime(40
K))-i(\Sigma^{\prime\prime}(7 K)-\Sigma^{\prime\prime}(40
K))]/\omega_{ph}(40K)$ and $u\equiv\pi+$
$2i\cosh^{-1}[\omega_{ph}(40K)/2\Delta]$. 
Using the energy of the strongest pair breaking peak $2\Delta$ =
135~\cm, we acquire weak e-p coupling constant
$\lambda_{B_{1g}}^\Gamma \approx 0.0015$. 
The \A\,\,phonons do not show measurable renormalization
(Figs.~\ref{Fig4}(d-e)). Therefore, the \B\,\,mode is the only
phonon that exhibits the SC induced self-energy effects and the total
e-p coupling constant $\lambda$ can be approximated by
$\lambda_{B_{1g}}^\Gamma$.

In comparison, for a conventional phonon-mediated superconductor
MgB$_2$ with similar \Tc\,\,= 39~K, a much larger coupling constant
$\lambda=0.2$ was 
derived from SC induced phonon
renormalization~\cite{Mialitsin_PRB2007}. 
Thus, for \CaK, the value $\lambda$ is far from being sufficient to
result in superconductivity with \Tc~at~35~K.

\textit{Conclusion}
-- In summary, we report polarization-resolved Raman spectroscopic
study of the single-crystal \CaK\,\,superconductor with \Tc\,\,at
35~K.

(1) We detect no signatures of Pomeranchuk-like electronic nematic 
fluctuations, which 
implies that the electronic nematicity may not be essential for
high-\Tc\,\,superconductivity in FeSCs. 

(2) Below \Tc, we observe
complete suppression of low-frequency spectral weight and development
of three SC pair-breaking peaks in the \BB\,\,symmetry channel, which
implies that all the SC gaps are nodeless. 
We do not detect in-gap collective modes which are commonly observed 
for FeSCs with strong electronic nematic fluctuations. 

(3) The pair-breaking peaks energies are about 20\% lower than twice
SC
gap energies measured by one-particle spectroscopy, likely due to 
finite-state interaction effects in the sub-dominant 
$d_{xy}$-symmetry pairing channel. 

(4) We study the SC induced self-energy effects for Raman-active
phonons
and provide an estimate of the electron-phonon coupling constant
$\lambda^\Gamma=0.0015$, which is very small for a superconductor
with \Tc\,\,at 35~K.

By showing that the electronic nematic fluctuations and
electron-phonon coupling are neglectfully small, we exclude them 
as a viable explanation for high-\Tc\,\,superconductivity in
\CaK. Naturally, one would consider the remaining spin-fluctuations 
to provide pairing interaction. 
In this case, the expected pairing symmetry is $s_\pm$, which is
consistent with
the observation of the spin-resonance mode at nesting vector
$(\pi,\pi)$ by inelastic neutron scattering~\cite{Iyo_JPSJ2018}.

Work at Rutgers was supported by the U.S. Department of Energy,
Office of Basic Energy Sciences, Division of Materials Sciences and
Engineering under Contract No. DE-SC0005463. 
Work at Ames Laboratory was supported by the U.S. Department of
Energy, Office of Basic Energy Sciences, Division of Materials
Sciences and Engineering under Contract No. DE-AC02-07CH11358.
W.R.M was supported by the Gordon and Betty Moore Foundation's EPiQS
Initiative through Grant No. GBMF4411.

\end{document}